# Heterogenous Dynamics in a Polymer Solution Revealed through Measurement of Ultraslow Convection

Thomas P. Chaney[1], Samuel D. Marks[1], Dylan M. Ladd[1], Andrei Fluerasu[2], Federico Zontone[3], Yuriy Chushkin[3], Sebastian Frücht[4,5], Dina Sheyfer[6], Kelsey Levine[1], Amnahir E. Peña-Alcántara[8], Hans-Georg Steinrück[4,5,7], Michael F. Toney[1,9]

*[1]Materials Science and Engineering, University of Colorado, Boulder, CO 80309, USA*
*[2]National Synchrotron Light Source II, Brookhaven National Laboratory, Upton, NY 11973-5000, United States*
*[3]ESRF - The European Synchrotron, 71 Avenue des Martyrs, CS 40220,38043, Grenoble Cedex 9, France*
*[4]Institute for a Sustainable Hydrogen Economy (INW), Forschungszentrum Jülich, Marie-Curie-Straße 5, 52428 Jülich, Germany*
*[5]Institute of Physical Chemistry, RWTH Aachen University, Landoltweg 2, 52074 Aachen, Germany*
*[6]Materials Science Division, Argonne National Laboratory, Argonne, Illinois 60439, USA*
*[7]Department of Chemistry, Paderborn University, Warburger Straße 100, 33098 Paderborn, Germany*
*[8]Department of Materials Science and Engineering, Stanford University, Stanford, CA 94305, USA*
*[9]Department of Chemical and Biological Engineering and Renewable and Sustainable Energy Institute (RASEI), University of Colorado, Boulder, CO 80309, USA*

**Abstract:**

Understanding solution-phase aggregation and dynamics in complex fluids is critical for material processing, yet widely used dynamic light scattering (DLS) fails for strongly attenuating systems such as conjugated polymers. We use X-ray photon correlation spectroscopy (XPCS) to probe the dynamics of a polymer, PM7, in toluene, revealing unexpected oscillations in the autocorrelation function that show vertical flow during measurement. Despite the relatively low X-ray absorption, measured flow velocities scale with X-ray beam power and suggest convective transport. Our analyses reveal both mobile and static scatterers that together produce oscillatory, heterodyne features in the measured correlation functions. Finite element simulations predict flow velocities much larger than observed, suggesting that entanglements of the aggregates slow their motion. These results provide a direct measurement of ultra-slow convection and highlight the need to explicitly account for even modest beam heating in interpreting XPCS results. Moreover, the observation of distinct scatterer populations underscores the structural complexity of conjugated polymer solutions.

**Background and Introduction:**

Particle size and motion in liquid-phase suspensions are of great importance towards understanding materials processing and function across a variety of fields including pharmaceuticals, polymers, and nanoparticles. Dynamic light scattering (DLS) is commonly used to probe the size and dynamics of nanoparticles, polymer aggregates, and proteins in solution. DLS measures size through the decay of the scattering-vector, $\vec{q}$, dependent time autocorrelation function of speckle intensities due to nanoscale dynamics. Measurements of various flow fields have been demonstrated through oscillations of the autocorrelation function [1,2]. A limitation of DLS is its non-applicability to solutions/particles that strongly absorb light and exhibit multiple scattering effects at the probing wavelength. These effects may reduce the scattered signal strength and introduce unintended dynamics [3]. Unintended dynamics  may arise from convective flows caused by local heating of the solution due to light absorption. Convective flow has been observed using DLS in numerous materials including conjugated polymers, hemoglobin, and gold nanoparticles [4–9]. The flow has also been imaged using tracer particles for visualization of the convective cycles [10]. Quantification of these convective flows has remained elusive due to the inability for most DLS experiments to accurately probe scattering with vertical $\hat{q}$ components [5]. Therefore, the nature of light absorption-induced convective currents and strategies to mitigate them remain poorly understood.

X-ray photon correlation spectroscopy (XPCS) presents new opportunities to better characterize flow dynamics due to the use of area detectors that simultaneously record scattering in both vertical and horizontal $\vec{q}$ directions spanning several decades in $q$ [11–16]. Also, X-rays have lower attenuation in most materials compared to visible light. Thus XPCS, in some cases, can overcome the limitations of DLS in measuring optically opaque solutions. In XPCS and DLS, the velocity of scatterers within a sample can be probed in homodyne or heterodyne modality [16–18]. Homodyne experiments probe the relative difference in velocity between scatterers, while heterodyne experiments resolve velocities through the use of a static reference scatterer [14,16].

In this study, we intended to use XPCS to probe the internal dynamics of aggregate structures present in a solution of a state-of-the-art conjugated polymer, (Poly[(2,6-(4,8-bis(5-(2-ethylhexyl)-4-chlorothiophen-2-yl)-benzo[1,2-b:4,5-b']dithiophene))-alt-(5,5-(1',3'-di-2-thienyl)-5',7'-bis(2-ethylhexyl)benzo[1',2'-c:4',5'-c']dithiophene-4,8-dione))]) referred to as PM7, in toluene. The chemical structure of PM7 is shown in **Figure S1**. By performing experiments with varying X-ray beam power, we identify unexpected oscillations in the autocorrelation functions that indicate vertical convective flow driven by local beam heating despite the relatively low attenuation of X-rays in the sample. Our analyses of these autocorrelation functions are consistent with the presence of both mobile and quasi-static polymer aggregates producing internal heterodyne mixing. Furthermore, the slower-than-expected flow velocities suggest the mobile polymer aggregates are likely constrained by entanglements leading to non-Newtonian flow behavior. Our results underscore the complex, entangled aggregate structures in solution-processed organic electronic materials and highlight the need to account for beam-induced heating in XPCS experiments on liquid samples.

**Methods:**

Speckle patterns in the small-angle X-ray scattering regime were collected from solutions of PM7 polymer in toluene solvent at the beamlines ID10 at ESRF and 11-ID at NSLS-II. PM7 was obtained from 1-Material Inc. with molecular weight of 120 kDa and polydispersity index of 2.5. In both experiments, the solution was prepared by dissolving 10 mg of PM7 per milliliter of toluene and stirring for at least 3 hours. These solutions were then transferred into a 1 mm diameter quartz capillary with 10 μm wall thickness, sealed, and mounted horizontally in a metal sample cell which was at ambient temperature (~20 °C). At ESRF ID10, a small beam size of 40 $\mu m^2$ was used producing an unattenuated flux density of $6.2x10^{10}$ photons/s/$\mu m^2$. Measurements were taken with three systematically varied upstream attenuators to achieve 47%, 9%, and 1% transmission of the unattenuated flux incident on the PM7 polymer solution with detector image collection times of 0.5, 0.5, and 1.0 s respectively. To rule out any contribution of experimental setup and further investigate the flux density dependance, experiments were repeated at NSLSII 11-ID with a much larger beam size of 1600 $\mu m^2$ producing an unattenuated flux density of $6.3x10^7$ photons/s/$\mu m^2$. Due to the lower flux density detector frames were collected at 5 s increments. For the remainder of the manuscript, we will refer to the two experimental setups as small-beam and large-beam for the ESRF ID10 and NSLS-II 11-ID experiments, respectively. Additional details on experimental setup are in **Tables S1-S2.** For both experimental setups, we ensured the longitudinal coherence condition is met for our maximum $q$ value of 0.0185 Å$^{-1}$ (**Figure S2** and **Table S3**). Single detector frames are shown for measurements taken at small-beam and large-beam conditions in **Figures 1a and 1d,** respectively. Data was analyzed using the beamlines' specific software [19]. For each set of detectors images, we computed two-time autocorrelation functions (TTCFs) spanning regions of the detector defined by a $q$-range and azimuthal angle, $\phi$ [20,21].

**Results and Discussion:**

The two-time correlation functions (TTCFs) for PM7 in toluene are plotted in **Figures 1b-c** for small-beam conditions with 47% of an unattenuated beam flux at $q$=0.005 Å$^{-1}$ at $\phi$=90° and $\phi$=0°. Oscillations appear at $\phi$=90° (**Figure 1b**) along both the constant-lag and the constant-age directions. These directions are shown in **Figure 1f** and their definition and meaning is described in detail elsewhere [20,21]. The checkerboard pattern in **Figure 1b** has been previously observed in epitaxial growth experiments [22,23] and for shear thickening colloids [24]. All oscillations disappear at $\phi$=0° (**Figure 1c**), indicating that they are caused by anisotropic dynamics such as directional flow [12,15]. The $\phi$-dependence of the TTCF features is apparent at a range of $q$-values and across all measurements as shown in **Figure S3-S5**. Importantly, the time-dependent small-angle scattering data shown in **Figure S6** shows no significant changes in scattering across the measurement time that would suggest beam-induced structural evolution. This observation casts significant doubt on the validity of the typical approach of using a constant scattering intensity (i.e., constant $S(q)$) as a proxy for the maximum dose at which intrinsic sample dynamics are still measurable (i.e., non-beam-altered $S(q, \omega)$) [25]. While noting that each sample system may behave differently with regard to “beam damage”, our results unambiguously show that a constant static *S(q)* is not a proxy, or even approximation thereof, for a constant dynamic *S(q, ω)* (even for cases when this changes along the lag time (**Figures S3-S5**)). In this context, **Figures S7-S8** confirm isotropy in the scattering pattern and consistency of the azimuthally integrated *I(q)* between measurements, respectively.

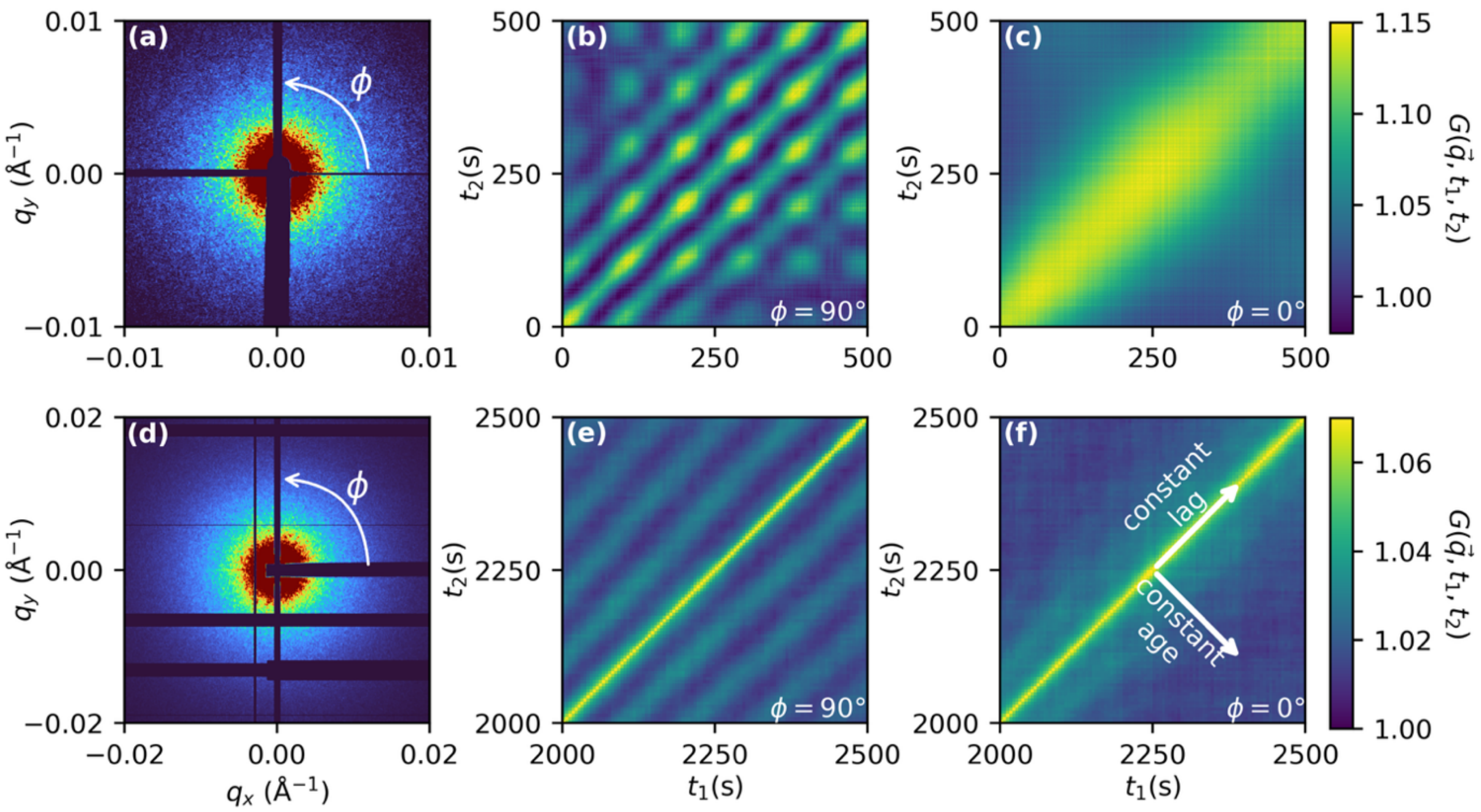

**Figure 1** (a) Linearly scaled detector frame from small-beam scattering at 47% beam flux with 0.5 s readout time. (b) TTCF of small-beam data taken at $q$=0.0045-0.0055 Å$^{-1}$ and $\phi$=90°. (c) TTCF of small-beam data taken at $q$=0.0045-0.0055 Å$^{-1}$ and $\phi$=0°. (d) Linearly scaled detector frame from large-beam scattering at 100% beam flux with a 5 s readout time. (e) TTCF of large-beam data taken at $q$=0.0185 Å$^{-1}$ and $\phi$=90°. (f) TTCF of large-beam data taken at $q$=0.0185 Å$^{-1}$ and $\phi$=0°. Note only a center portion of the TTCF is shown in (e) and (f) to allow for easy comparison with (b) and (c) as well as to avoid bad frames. Full TTCFs are shown in **Figures S3-S5**.

To understand the oscillations in autocorrelation value along the constant-lag direction (**Figure 1b**), we analyze the large-beam data shown in **Figures 1e-f**. The TTCF shows no oscillations along the constant-lag direction but still exhibits a similar oscillation trend along the constant-age diagonal with varying $\phi$, compared to the small-beam data. Given the ~40x greater beam area in the large-beam experiments, we suggest the oscillations along the constant-lag direction seen in the small-beam data (**Figure 1b**) are likely due to the X-ray probe capturing local dynamics that produce periodically increased speckle contrast which averages out with a larger beam cross-section. Indeed, waterfall plots in **Figure S9** show concurrent speckle appearance and disappearance in the small-beam data that accounts for the changes in contrast. The small beam cross section captures the local motion of a single or small number of aggregates which would produce concurrent, cyclical fluctuations of speckles from the heterodyne contrast described below. We can roughly estimate the number of aggregates within the beam volume through the polymer volume fraction of ~1% in the solution. Using an estimated cylindrical aggregate diameter of 100 Å from previous studies [26], and length of 1500 Å corresponding to the average molecular weight of the PM7, we estimate ~23 million aggregates within the small-beam volume. However, previous SAXS studies suggest significant polydispersity in aggregate size [26], and our XPCS is only probing $q$-ranges corresponding to aggregate diameters >500 Å. Additionally, each slice of the X-ray detector only encompasses a small portion of the orientation sphere for our isotropic liquid sample. For these reasons it is plausible that only a relatively small number of individual aggregates are producing speckles captured in the small-beam TTCFs. A similar strategy employing a focused beam was recently used to reveal similar local motion of scatterers through heterodyne behavior in shear-thickening colloidal solutions [24].

To gain further insight into the origin of the heterodyne-like oscillations in the TTCFs, we model the $g_2$ functions, obtained by time averaging the TTCF along the "constant-lag" direction as shown in **Figure 1f** [20,21]. The $g_2$ functions averaged over the center 50 s of the total scan time from a range of azimuthal angles are plotted in **Figure 2**. The oscillation frequency in the $g_2$ functions depend on the $q$-bin, as shown in **Figure S10**, consistent with previous XPCS experiments on systems with flow [16,24]. We first attempted to use a homodyne model which was unable to fit the data (**Figure S11)** [13]. Instead we found the experimental $g_2$ functions are modeled well by a heterodyne system characterized by mobile and static scattering populations represented by equation 1 [27,28].

$$g_2(\vec{q},\tau) = 1 + \beta\left(X^2\left|g_{1,A}(\vec{q},\tau)\right|^2 + (1-X)^2\left|g_{1,B}(\vec{q},\tau)\right|^2 + 2X(1-X)Re\left[g_{1,A}(\vec{q},\tau)\bullet g_{1,B}^{*}(\vec{q},\tau)\right]\right) \quad (1)$$

In equation 1, $\beta$ is the contrast, $X$ is the fraction of scattering from the mobile scatterers, and $g_{1,A}$ along with $g_{1,B}$ correspond to the intermediate scattering functions of mobile and quasi-static scatterers, respectively given by

$$g_{1,A}(\vec{q},\tau) = e^{i\tau\vec{q}\bullet\vec{v}}e^{-(\Gamma_{\mathrm{A}}\tau)^{\alpha_A}} \quad (2)$$

$$g_{1,B}(\vec{q},\tau) = e^{-(\Gamma_{\mathrm{B}}\tau)^{\alpha_B}} \quad (3)$$

Oscillations only appear in the TTCFs extracted from scattering bins that contain a vertical component of q, showing that the direction of the flow is vertical. Values of $|\vec{v}|$ from the fits shown in **Figure 2** indicate flow with velocities ranging from 3-13 Å/s depending on X-ray beam flux. Complete fit parameters are available in **Tables S4-S7**. We note that this model is likely a simplification of the actual system which that has large variations of $\Gamma_{\mathrm{A}}$, $\Gamma_{\mathrm{B}}$, $\alpha_A$, and $\alpha_B$. Nonetheless, the strong heterodyne oscillations allow reliable extraction of velocities and indicate that there are two populations of scatterers–one mobile and one quasi-static. These two populations can be explained through the presence of both mobile polymer aggregates and a static polymer network of aggregates. Previous SAXS and cryo-EM studies of PM7 in chlorobenzene solutions support this explanation suggesting the presence of a complex population of aggregates exhibiting both isolated and networked structures [26,29].

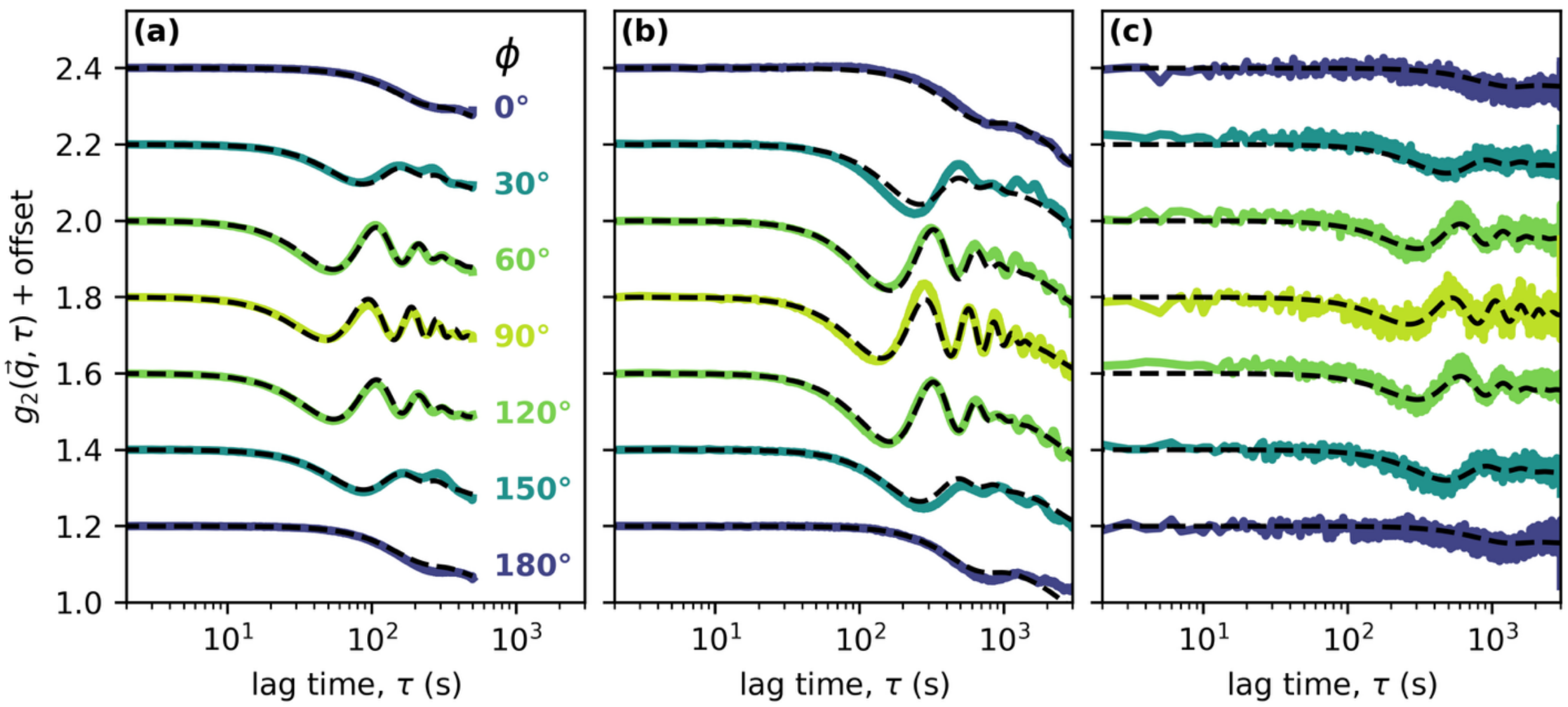

**Figure 2** One-time correlation ($g_2$) functions (solid lines) and fits (dashed lines) for small-beam data obtained using an average over the center 50 s of the total scan time of the TTCFs. Cuts shown are obtained from TTCFs evaluated at various $\phi$ values with width of 30° and a $q$-range of 0.0045-0.0055 Å$^{-1}$. Fits from three beam intensities are shown (a) 47%, (b) 9%, and (c) 1% of unattenuated flux. Curves are vertically offset by 0.2 for clarity.

Having established the origin of the internal heterodyne mixing, we now discuss the driving force for the vertical motion of the mobile aggregates. In the XPCS experiment, the horizontally mounted capillary is sealed on both ends and no external fields were applied. Thus, the vertical motion must be due to gravity–the only force acting on the aggregates in the vertical direction. Gravity can drive vertical flow of aggregates through sedimentation or buoyancy due to a density difference with the surrounding matrix.

To understand whether sedimentation or buoyancy is driving the vertical velocity, we examine the dependence of fitted velocity on X-ray beam flux. The vertical velocities extracted from the fits in **Figure 2** are plotted versus the absorbed X-ray power density in **Figure 3** revealing a linear dependence. While X-ray absorption can have many effects on a sample, the lack of major structural rearrangements, indicated by the unchanged SAXS profile and intensity and the identification of gravity as the driving force, suggest local heating due to X-ray absorption is the main factor.

Local heating of a fluid decreases the dynamic viscosity, $\eta$, and solvent density, $\rho_{\text{solvent}}$, while the particle density $\rho_{\text{particle}}$ remains the same. Therefore, the sedimentation speed, $v$, will increase under the acceleration of gravity, $g$, as approximated through the Stokes equation:

$$v = \frac{2gr^2(\rho_{\text{particle}} - \rho_{\text{solvent}})}{9\eta} \quad (4)$$

However, using an approximate aggregate radius, $r$ of 100 Å as found in previous solution SAXS studies [26], and the tabulated temperature dependent viscosity and density of toluene, we find the temperature of the solution within the beam volume would need to reach ~210 °C for sedimentation to explain the velocities, an unrealistic temperature higher than the boiling point of toluene (**Figure S12**).

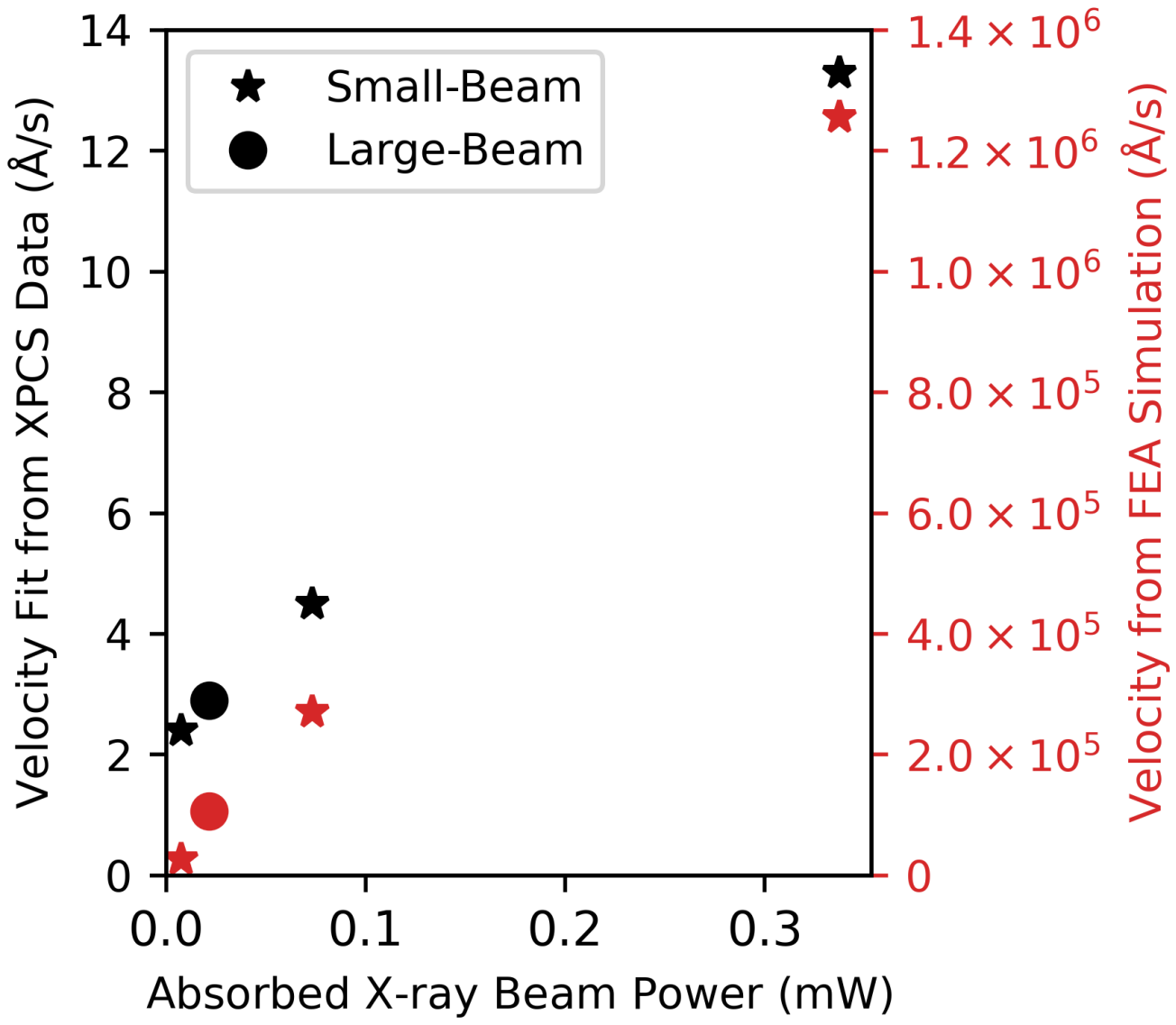

**Figure 3** Comparison of velocities from XPCS analysis and FEA simulation plotted against X-ray beam power.

We are now left with convective currents driven by the buoyant locally heated solvent dragging aggregates upward as the only reasonable mechanism driving the vertical velocity.

Indeed, this effect has been identified as a driving force for beam-induced vertical flow in DLS experiments [4–8]. However, most DLS experiments are unable to accurately measure the velocities due to uncertainty in the vertical scattering vector caused by thermal lensing [5]. Our XPCS measurements present the opportunity to directly compare velocities fitted from autocorrelation functions to predicted velocities via fluid dynamics models.

To find the fluid velocities that can be expected from beam induced convection, we perform finite element analysis (FEA) of a two-dimensional representation of our experiment using COMSOL Multiphysics software. The X-ray beam absorption is modeled as a 2D Gaussian heat source and the capillary walls held at a temperature of 20 °C. In our FEA simulations we utilize material properties of neat toluene since viscosity measurements of the PM7 solution, as shown in **Figure S13**, indicate little deviation from the behavior of pure toluene. Full details of the FEA simulations are included in the supplemental information (**Table S8**), and an example of flow field and temperature distribution from a steady-state solution for the XPCS experiment are shown **Figures S14-S15**. We compare the vertical velocity in the beam volume from the simulations with results from XPCS analysis in **Figure 3**. While the velocities predicted from FEA also increase monotonically with beam power, they are surprisingly several orders of magnitude faster than velocities obtained from XPCS autocorrelation functions.

One explanation may be the competing sedimentation velocity of these particles in the opposite direction. However, as shown in **Figure S12** we expect the sedimentation velocity for ~100 Å aggregates to be on the order of Å/s, not nearly enough to impact the vertical convective flows of µm/s as estimated from FEA simulations. It is also possible that the previously assumed static population of aggregates is also mobile. However, this is also unlikely that we have a purely bimodal distribution of aggregate velocities giving rise to strong heterodyne features. Instead, we propose that the viscosity of the solution at these very low shear rates is much larger than the viscosity of pure toluene. Our viscosity measurements in **Figure S13** were taken at a shear rate of ~20,000 $s^{-1}$, several orders-of-magnitude higher than the estimated shear in solution of <1 $s^{-1}$. Polymer solutions are known to exhibit a non-Newtonian shear-thinning behavior where at very small shear rates a solution may have orders-of-magnitude greater viscosities caused by weak entanglements that provide resistance to flow at low shears but that are easily broken at higher shears [30]. A study of a similar conjugated polymer in solution with chlorobenzene found that dilute solutions exhibit low-shear rate viscosities of over 10 Pa•s, four orders of magnitude larger than the viscosity of the neat solvent [31]. Using a value of 10 Pa•s for the viscosity produces flow rates on the order of 10 Å/s in agreement with the XPCS data. We suggest the drastic non-Newtonian behavior of conjugated polymer solution is responsible for the observed ultra-slow convective velocities.

In this work, XPCS has been used to reveal the presence of both static and mobile aggregate structures in a PM7 conjugated polymer solution with toluene. By using X-rays instead of visible light, we circumvent the effects of strong absorption on the solution dynamics. We found that the local heating from the X-ray beam absorption produces convective currents with velocity proportional to beam power. This phenomenon has been previously observed but not well-quantified in DLS experiments. Through our analyses of the autocorrelation function obtained at various azimuthal angles on the area detector, we reveal isolated vertical flow with velocities on the order of Å/s. These are orders of magnitude slower than predicted by simple FEA simulations. We attribute the discrepancy in velocity to entangled aggregate structures that introduce shear-thinning behavior resulting in viscosities at low shear rates of ~10 Pa•s. Overall, this study underscores the complexity of conjugated polymer solution aggregates that can impact the morphology and performance of solution deposited thin films. Additionally, we demonstrate that a constant scattering intensity profile (e.g. constant $S(q)$) is not sufficient to assume unchanging sample dynamics or the lack of beam driven effects within the sample. Notably, we characterize X-ray induced convection that is likely present in many solution-phase X-ray scattering and light scattering experiments–an effect which must be carefully considered for design of future solution phase XPCS and DLS experiments.

**Data Availability:**

X-ray photon correlation spectroscopy detector images, two-time correlation functions, analysis workflows, and COMSOL files are openly available at DOI:10.5281/zenodo.17207775.

**Acknowledgments:**

This work was supported by the Office of Naval Research MURI Program Center for Self-Assembled Organic Electronics (SOE), Grant N00014-19-1-2453. T.P.C. acknowledges support from the National Science Foundation Graduate Research Fellowship Program under Grant No. (DGE 2040434). A portion of the XPCS characterization was conducted at beamline 11-ID of the National Synchrotron Light Source II, a U.S. Department of Energy (DOE) Office of Science User Facility operated for the DOE Office of Science by Brookhaven National Laboratory under contract no. DE-SC0012704. We also acknowledge the European Synchrotron Radiation Facility (ESRF) for provision of synchrotron radiation facilities for another portion of XPCS characterization under proposal number MA-5746 using beamline ID10. DML acknowledges support from the National Science Foundation Center for Integration of Modern Optoelectronic Materials on Demand (IMOD) under Cooperative Agreement No. DMR-2019444 and from the Graduate Research Fellowship Program (NSF-GRFP) under Grant No. DGE 2040434. HGS and SF acknowledge the funding by the German Federal Ministry of Research, Technology and Space (BMFTR) and the Ministry of Economic Affairs, Industry, Climate Action and Energy of the State of North Rhine-Westphalia through the project HC-H2, and from the BMBF via projects 05K22PP2 and 05K24CJ2. A.PA. acknowledges support from the National Science Foundation Graduate Research Fellowship Program under Grant No. DGE-1656518, as well as the Stanford Knight-Hennessy Scholarship and the Stanford Enhancing Diversity in Graduate Education Doctoral Fellowship. We are grateful for Ankur Gupta's assistance and fruitful discussions on fluid dynamics calculations.

## Supplemental Information for:

# Heterogenous Dynamics in a Polymer Solution Revealed through Measurement of Ultraslow Convection

Thomas P. Chaney[1], Samuel D. Marks[1], Dylan M. Ladd[1], Andrei Fluerasu[2], Federico Zontone[3], Yuriy Chushkin[3], Sebastian Frücht[4,5], Dina Sheyfer[6], Kelsey Levine[1], Amnahir E. Peña-Alcántara[8], Hans-Georg Steinrück[4,5,7], Michael F. Toney[1,9]

*[1]Materials Science and Engineering, University of Colorado, Boulder, CO 80309, USA*
*[2]National Synchrotron Light Source II, Brookhaven National Laboratory, Upton, NY 11973-5000, United States*
*[3]ESRF - The European Synchrotron, 71 Avenue des Martyrs, CS 40220,38043, Grenoble Cedex 9, France*
*[4]Institute for a Sustainable Hydrogen Economy (INW), Forschungszentrum Jülich, Marie-Curie-Straße 5, 52428 Jülich, Germany*
*[5]Institute of Physical Chemistry, RWTH Aachen University, Landoltweg 2, 52074 Aachen, Germany*
*[6]Materials Science Division, Argonne National Laboratory, Argonne, Illinois 60439, USA*
*[7]Department of Chemistry, Paderborn University, Warburger Straße 100, 33098 Paderborn, Germany*
*[8]Department of Materials Science and Engineering, Stanford University, Stanford, CA 94305, USA*
*[9]Department of Chemical and Biological Engineering and Renewable and Sustainable Energy Institute (RASEI), University of Colorado, Boulder, CO 80309, USA*

**Figure S1**: Chemical Structure of PM7 (Poly[(2,6-(4,8-bis(5-(2-ethylhexyl)-4-chlorothiophen-2-yl)-benzo[1,2-b:4,5-b']dithiophene))-alt-(5,5-(1',3'-di-2-thienyl)-5',7'-bis(2-ethylhexyl)benzo[1',2'-c:4',5'-c']dithiophene-4,8-dione))]).

**Table S1:** Experimental Parameters for XPCS experiments at NSLS-II 11-ID (large-beam) and ESRF ID10 (small-beam).

| Facility | Photon Energy (keV) | Beam Size ($\mu m^2$) | Unattenuated Flux (ph/s) | Unattenuated Flux Density (ph/s/$\mu m^2$) | Detector Distance (m) |
|---|---|---|---|---|---|
| NSLS-II 11-ID | 12 | 40 x 40 | ~1.0 x $10^{11}$ | ~6.3 x $10^{7}$ | 10.08 |
| ESRF ID10 | 9.94 | 8.7 x 4.6 | ~2.5 x $10^{12}$ | ~6.2 x $10^{10}$ | 6.98 |

**Table S2:** Detector details for XPCS experiments at NSLS-II 11-ID (large-beam) and ESRF ID10 (small-beam).

| Facility | Detector Model | Pixel Size ($\mu m^2$) | Maximum Frame Rate (kHz) |
|---|---|---|---|
| NSLS-II 11-ID | Eiger 4M | 75 x 75 | 0.75 |
| ESRF ID10 | Eiger 500k | 75 x 75 | 22 |

**Longitudinal coherence condition:**

Change in path length, $\Delta L$, from scattering at front of sample and back of sample is calculated in equation S1. $\Delta L$ must be less than the longitudinal coherence length as shown in equation S2 [1].

$$\Delta L = \frac{w(1-\cos(2\theta))}{\cos(2\theta)} \approx 2w\theta^2 \qquad (S1)$$

$$2w\theta^2 < \frac{\lambda^2}{(2\Delta\lambda)} \qquad (S2)$$

$\lambda$ is the wavelength, $\Delta\lambda$ is the wavelength dispersion, $w$ is the sample width, $2\theta$ is the scattering angle and $\Delta L = L_1 - L_2$ as defined in **Figure S2**.

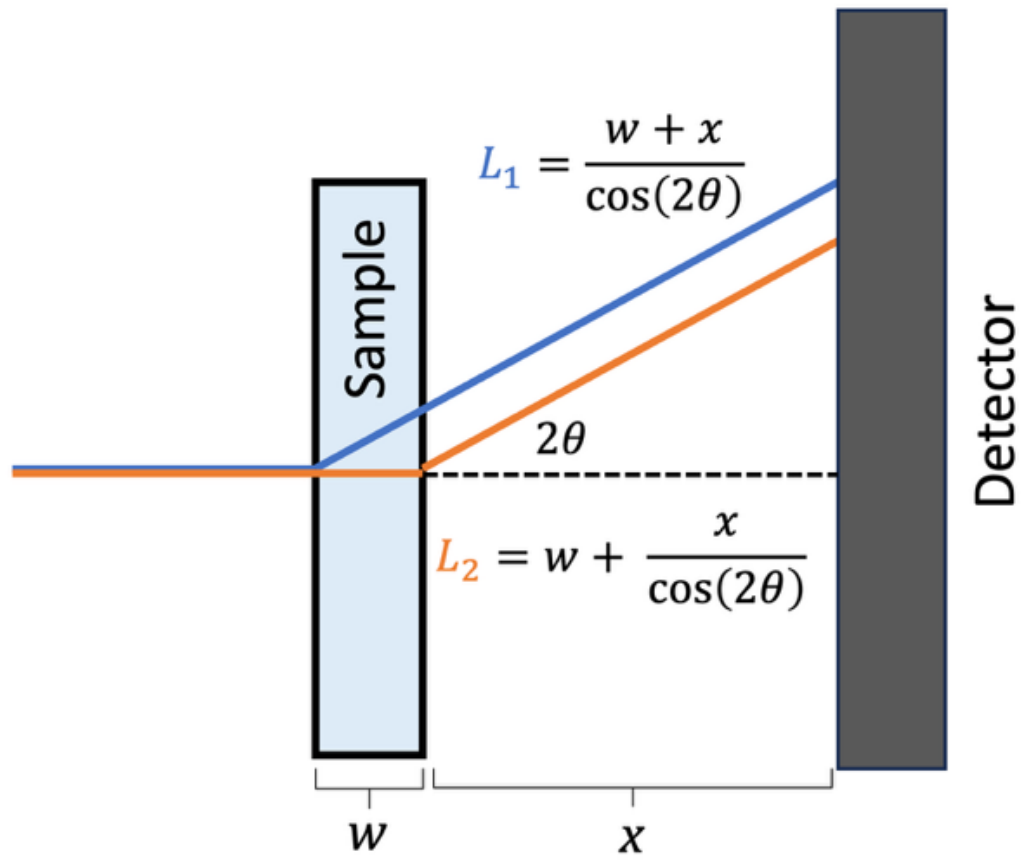

**Figure S2**: Diagram of scattering geometry dimensions used in equations S1 and S2.

**Table S3:** Using parameters from both beamlines used we calculate the maximum scattering vector $q$ that satisfies longitudinal coherence across the thickness of our sample.

| Facility | Wavelength (Å) | Dispersion (Å) | Width (mm) | $\theta_{max}$ (rad) | $q_{max}$ (Å$^{-1}$) |
|---|---|---|---|---|---|
| NSLS-II 11-ID | 1.033 | 0.00100 | 1.0 | 0.0016 | 0.020 |
| ESRF ID10 | 1.248 | 0.00035 | 1.0 | 0.0033 | 0.034 |

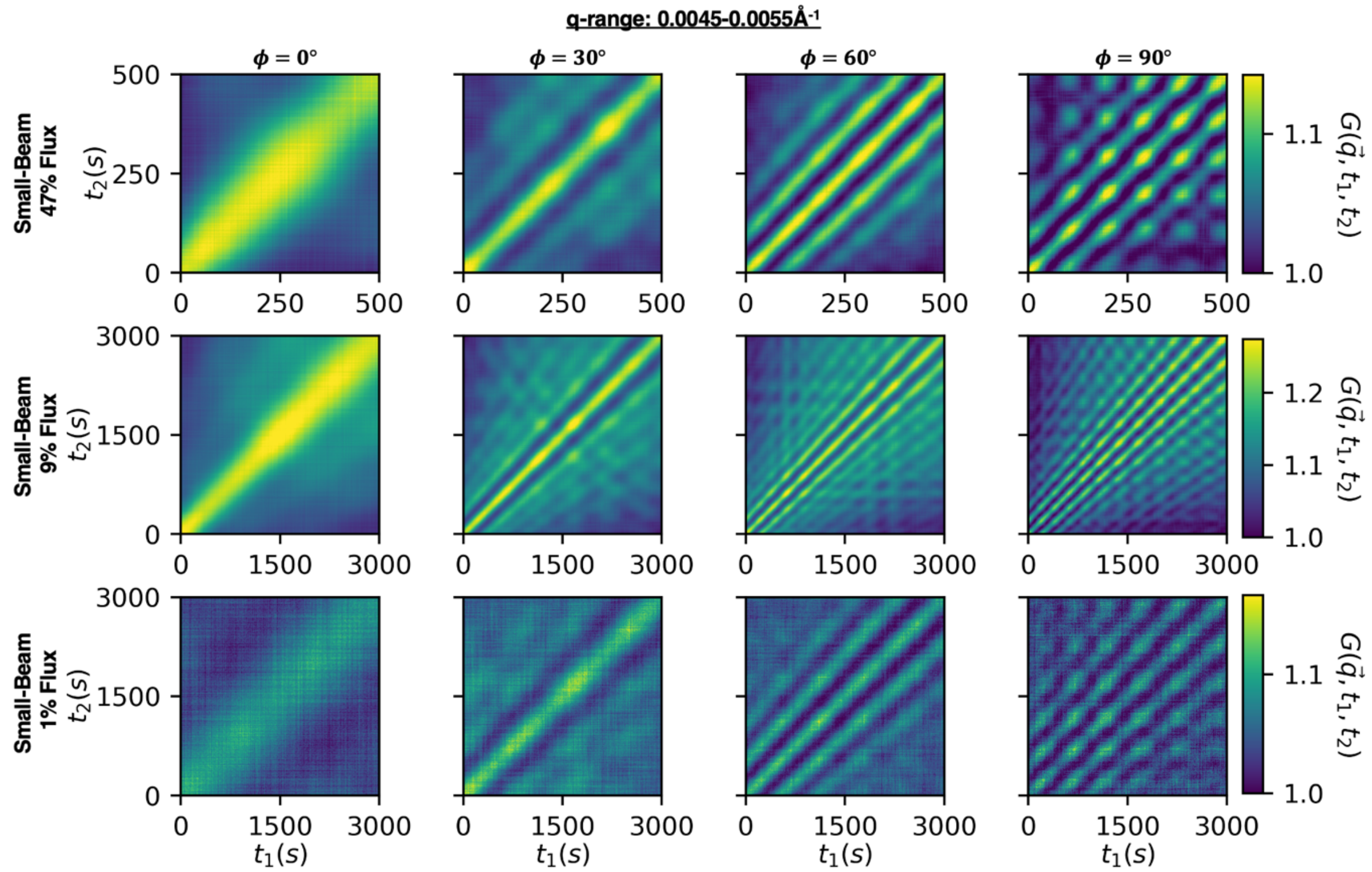

**Figure S3:** TTCFs at different $\phi$ values and fluxes for small-beam measurements at $q$=0.005Å$^{-1}$

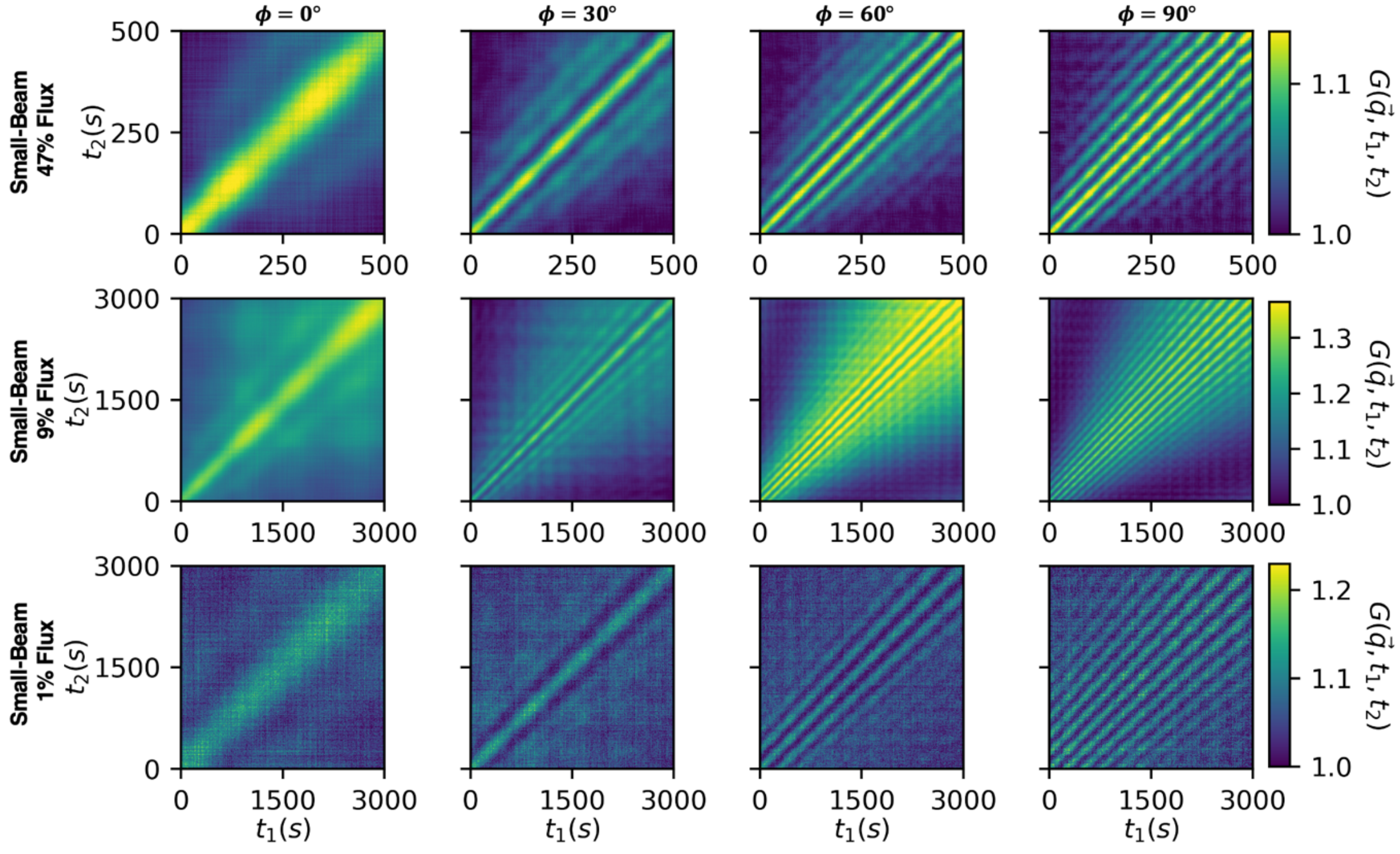

**Figure S4:** TTCFs at different $\phi$ values and fluxes for small-beam measurements at $q$=0.009Å$^{-1}$

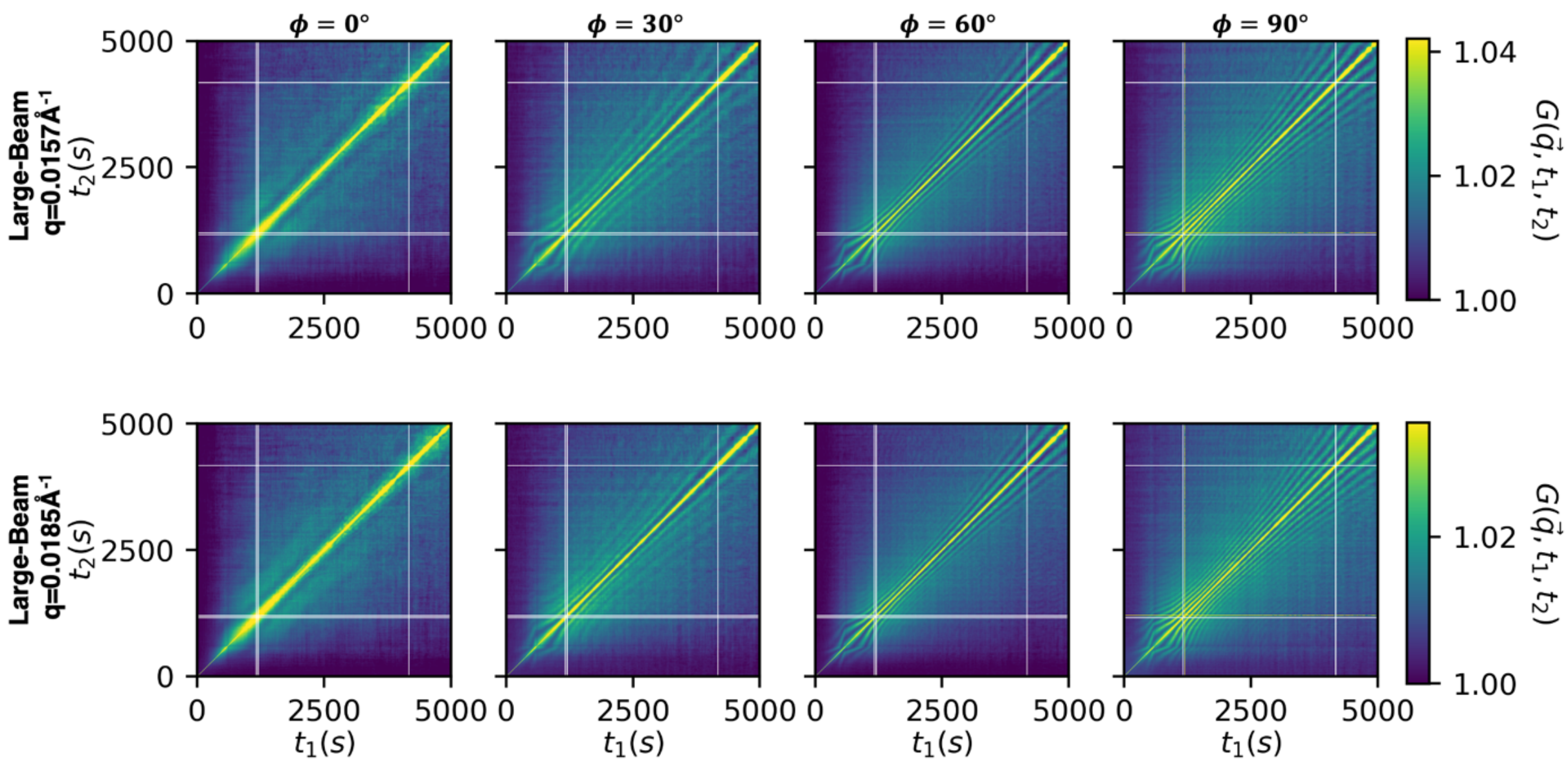

**Figure S5:** TTCFs at different $\phi$and $q$-values for large-beam experiments at NSLS-II. Note the white lines are due to bad detector frames.

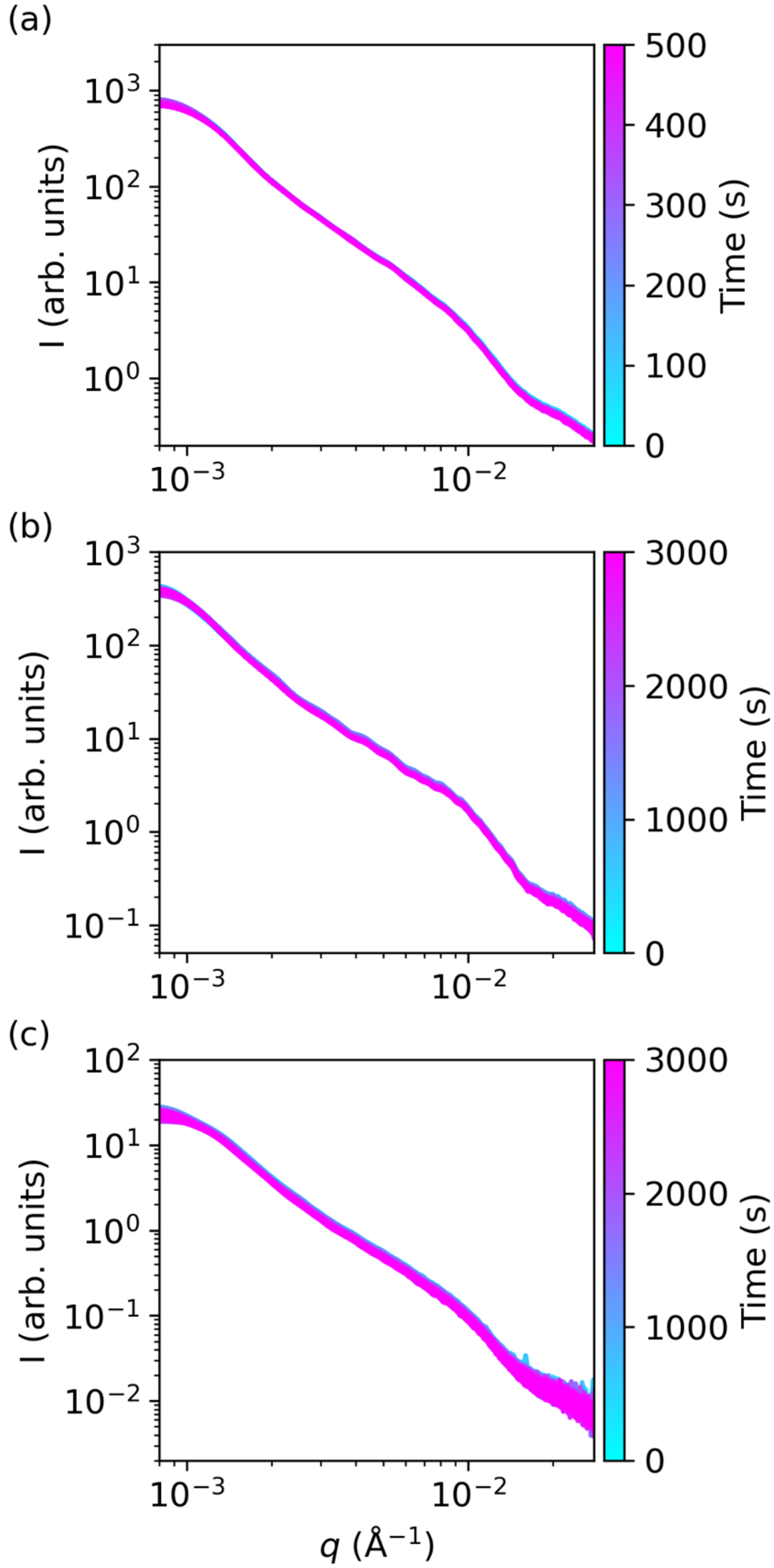

**Figure S6:** Azimuthally integrated small angle scattering data from each frame showing lack of structural damage and no significant intensity change for experiments conducted with the small-beam experimental setup with (a) 47% flux, (b) 9% flux, and (c) 1% flux

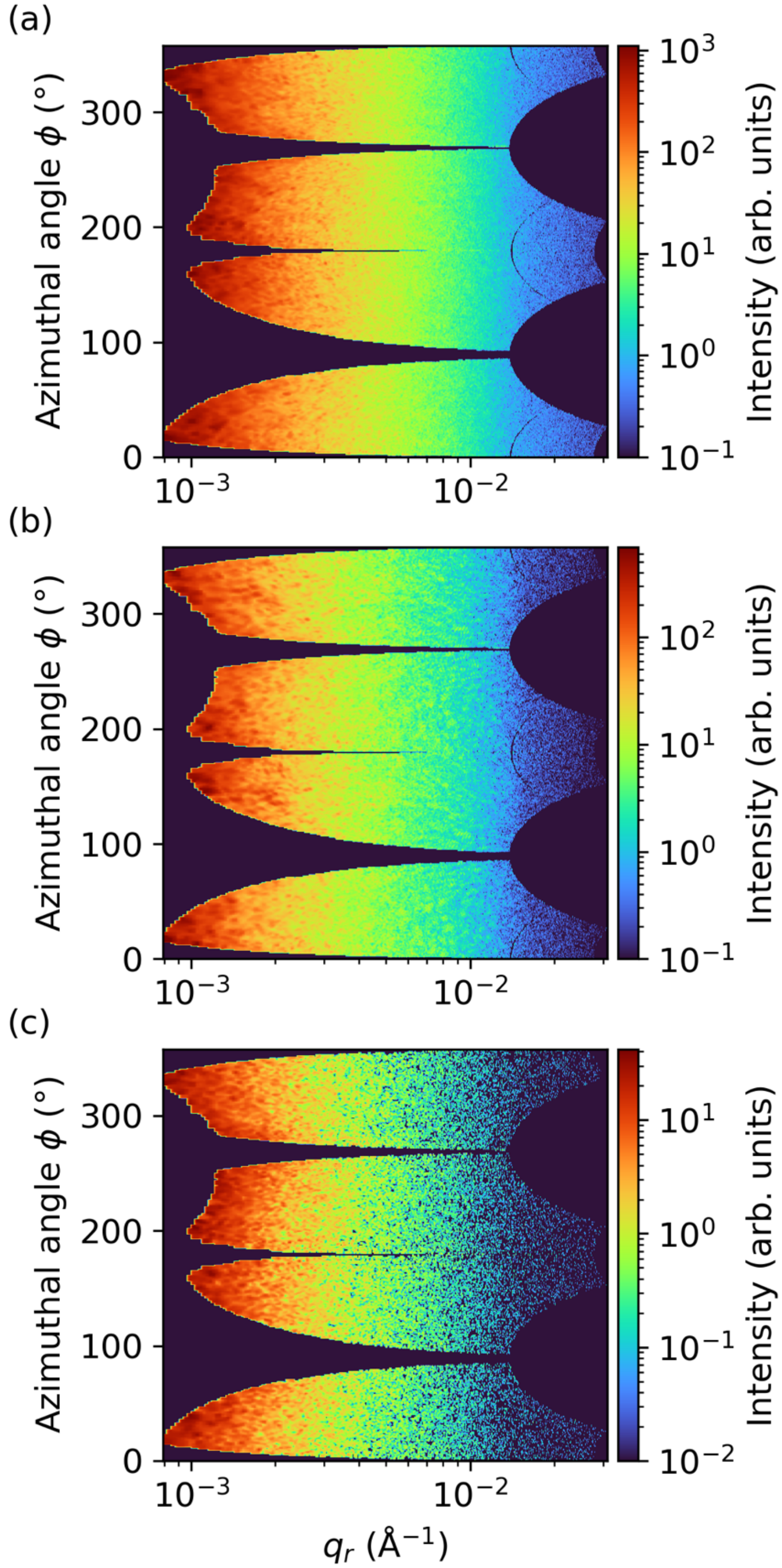

**Figure S7:** Small angle scattering polar plots showing isotropy in scattering profiles for experiments conducted with the small-beam with (a) 47% flux, (b) 9% flux, and (c) 1% flux

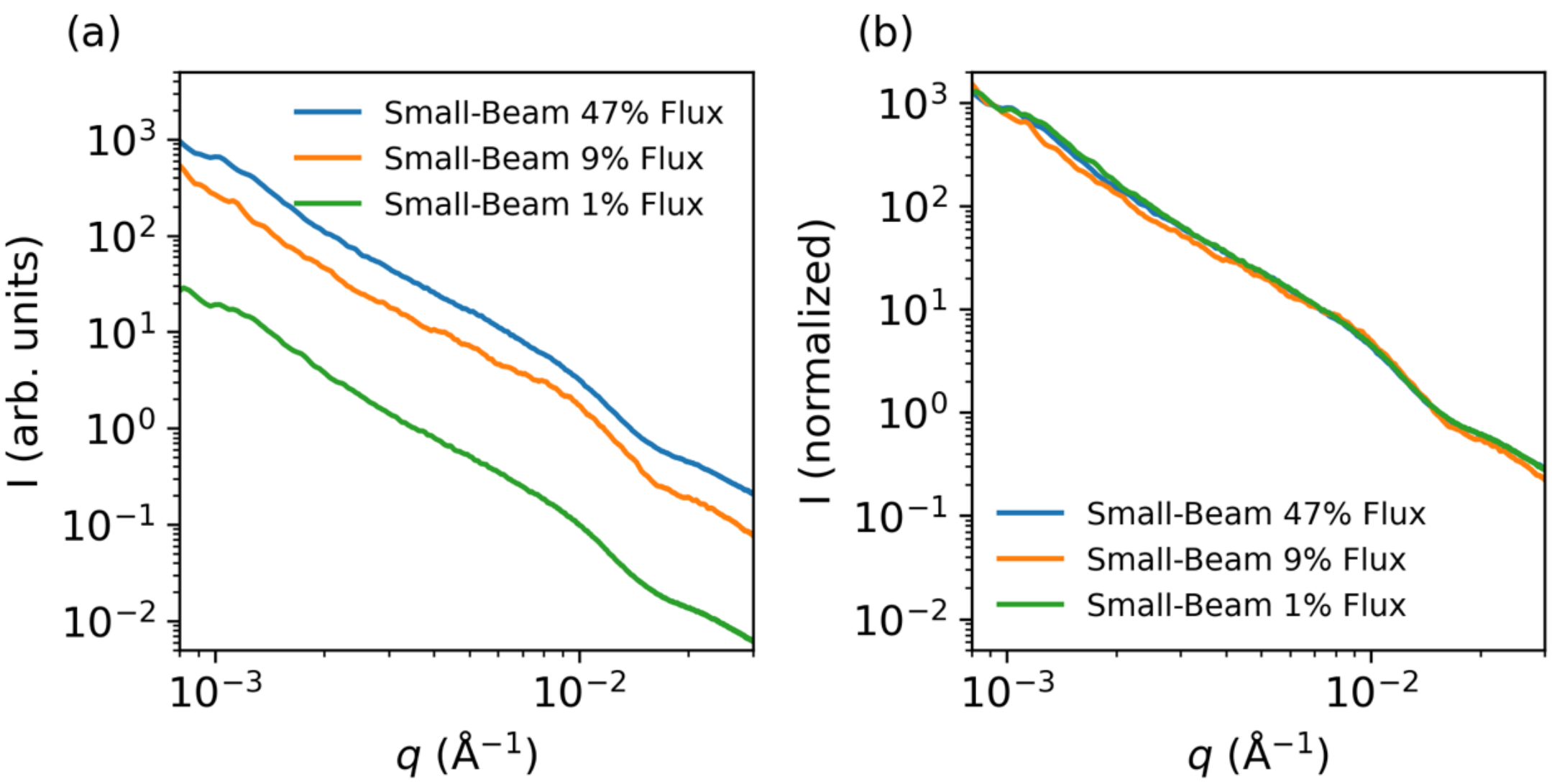

**Figure S8:** Time averaged and azimuthally integrated small angle scattering showing similar scattering profiles for each measurement. (a) unscaled data at the different X-ray flux, (b) the scaled profiles showing overlap.

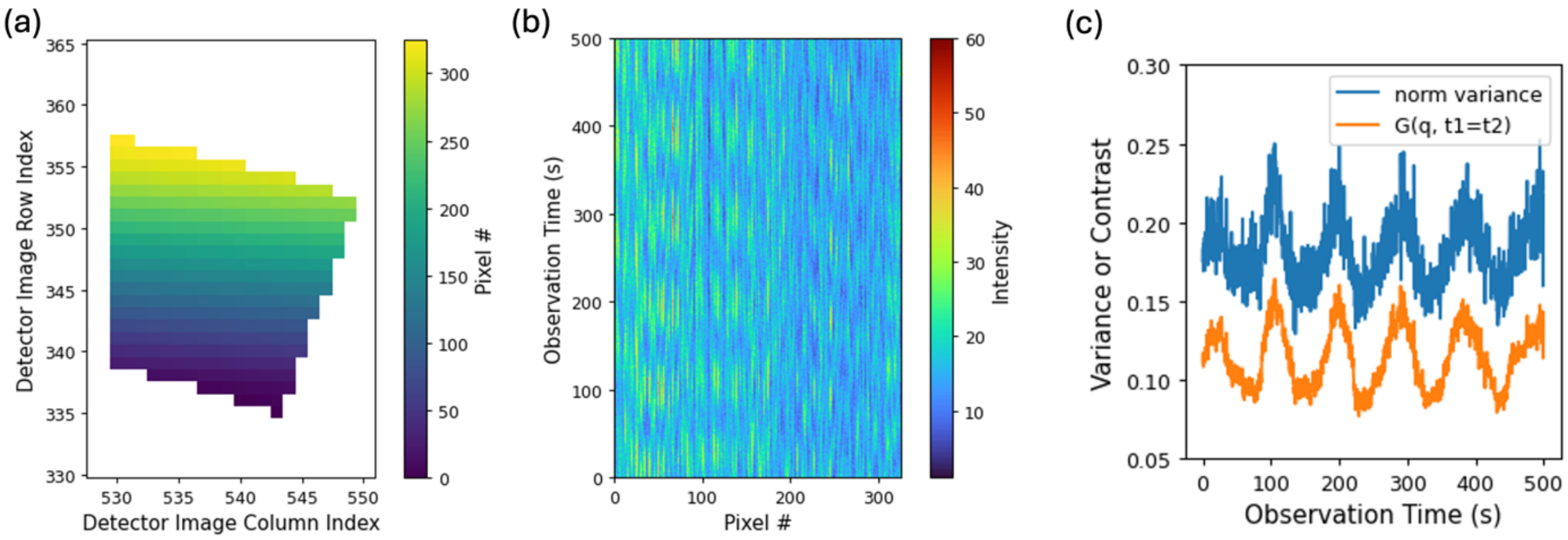

**Figure S9:** (a) plot detailing assignment of a slice of detector pixels to pixel numbers. (b) waterfall plots showing the change in each pixel intensity over time for the small-beam experiment at 47% flux. Notably, many pixels show concurrent oscillations in intensity. (c) plot showing oscillations along constant lag time diagonal (orange) are due to changes in the pixel intensity variance (blue).

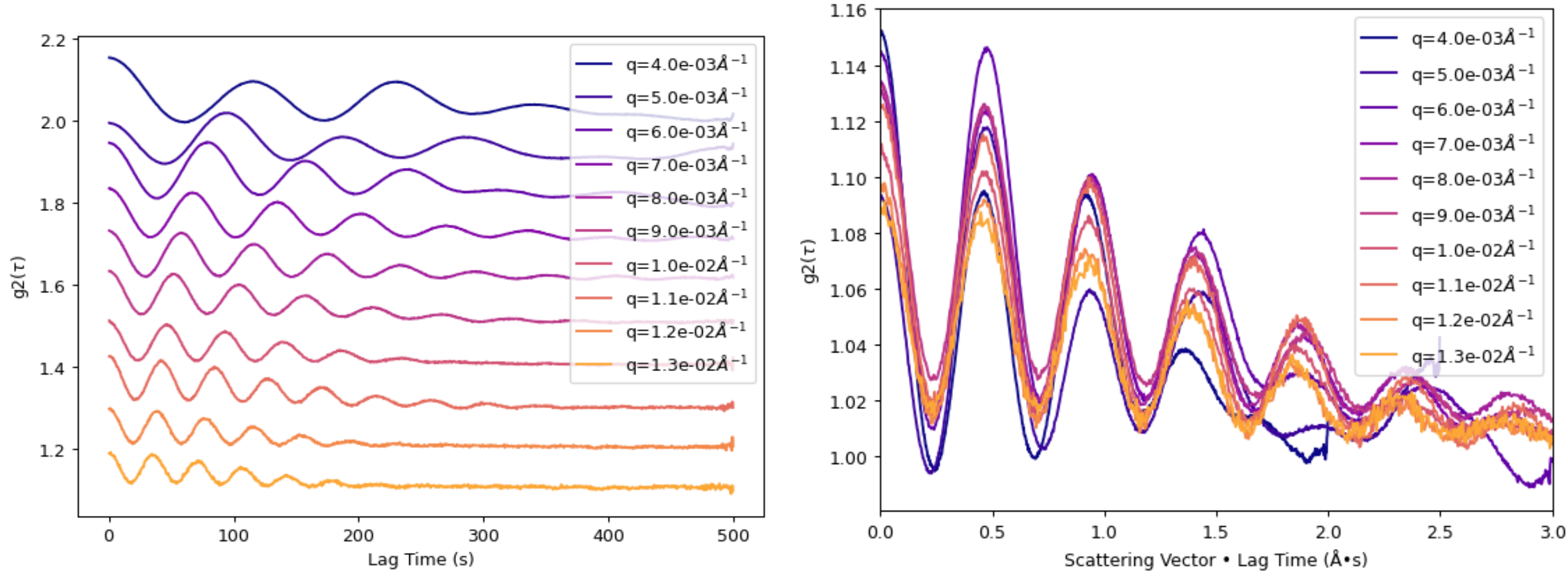

**Figure S10:** (left) demonstration of ripple frequency scaling with q, and (right) an overlay showing agreement in the probed velocity through a rescaled x-axis. Both plots are taken from the small-beam 47% flux experiment data at $\phi$=90°.

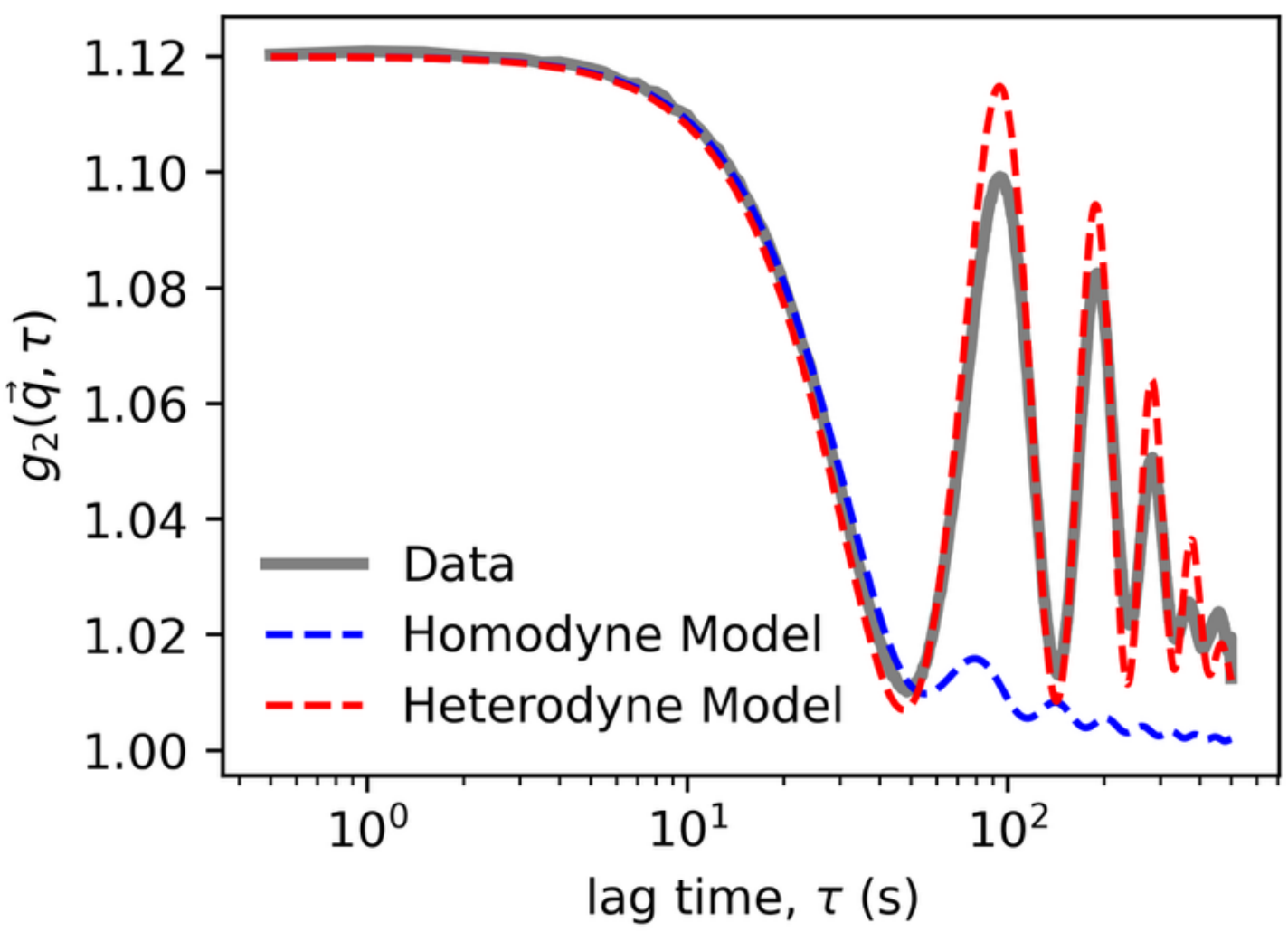

**Figure S11:** Demonstration of the inability for homodyne Poiseuille flow model to fit the experimental $g_2$ function from the small-beam at 47% flux, $q$=0.005Å$^{-1}$, and $\phi$=90°.

**Table S4:** Fitting Parameters for ESRF (small-beam) 47% flux. All $g_2$ functions at different $\phi$ values were co-fit allowing only $\beta$ to independently vary.

| $\phi$ (°) | β | X | q (Å$^{-1}$) | v (Å/s) | $\Gamma_A$ (s$^{-1}$) | $\alpha_A$ | $\Gamma_B$ (s$^{-1}$) | $\alpha_B$ |
|---|---|---|---|---|---|---|---|---|
| 0 | 0.14 | 0.62 | 0.005 | 13.4 | 0.0017 | 2.5 | 0.0029 | 2.5 |
| 30 | 0.13 | 0.62 | 0.005 | 13.4 | 0.0017 | 2.5 | 0.0029 | 2.5 |
| 60 | 0.14 | 0.62 | 0.005 | 13.4 | 0.0017 | 2.5 | 0.0029 | 2.5 |
| 90 | 0.12 | 0.62 | 0.005 | 13.4 | 0.0017 | 2.5 | 0.0029 | 2.5 |
| 120 | 0.13 | 0.62 | 0.005 | 13.4 | 0.0017 | 2.5 | 0.0029 | 2.5 |
| 150 | 0.13 | 0.62 | 0.005 | 13.4 | 0.0017 | 2.5 | 0.0029 | 2.5 |
| 180 | 0.15 | 0.62 | 0.005 | 13.4 | 0.0017 | 2.5 | 0.0029 | 2.5 |

**Table S5:** Fitting Parameters for ESRF (small-beam) 9% flux. All $g_2$ functions at different $\phi$ values were co-fit allowing only $\beta$ to independently vary.

| $\phi$ (°) | β | X | q (Å$^{-1}$) | v (Å/s) | $\Gamma_A$ (s$^{-1}$) | $\alpha_A$ | $\Gamma_B$ (s$^{-1}$) | $\alpha_B$ |
|---|---|---|---|---|---|---|---|---|
| 0 | 0.29 | 0.26 | 0.005 | 4.5 | 0.0011 | 2.5 | 0.00031 | 2.5 |
| 30 | 0.24 | 0.26 | 0.005 | 4.5 | 0.0011 | 2.5 | 0.00031 | 2.5 |
| 60 | 0.24 | 0.26 | 0.005 | 4.5 | 0.0011 | 2.5 | 0.00031 | 2.5 |
| 90 | 0.21 | 0.26 | 0.005 | 4.5 | 0.0011 | 2.5 | 0.00031 | 2.5 |
| 120 | 0.24 | 0.26 | 0.005 | 4.5 | 0.0011 | 2.5 | 0.00031 | 2.5 |
| 150 | 0.21 | 0.26 | 0.005 | 4.5 | 0.0011 | 2.5 | 0.00031 | 2.5 |
| 180 | 0.24 | 0.26 | 0.005 | 4.5 | 0.0011 | 2.5 | 0.00031 | 2.5 |

**Table S6:** Fitting Parameters for ESRF (small-beam) 1% flux. All $g_2$ functions at different $\phi$ values were co-fit allowing only $\beta$ to independently vary.

| $\phi$ (°) | β | X | q (Å$^{-1}$) | v (Å/s) | $\Gamma_A$ (s$^{-1}$) | $\alpha_A$ | $\Gamma_B$ (s$^{-1}$) | $\alpha_B$ |
|---|---|---|---|---|---|---|---|---|
| 0 | 0.09 | 0.30 | 0.005 | 2.4 | 0.00051 | 2.5 | 0.00006 | 2.5 |
| 30 | 0.11 | 0.30 | 0.005 | 2.4 | 0.00051 | 2.5 | 0.00006 | 2.5 |
| 60 | 0.09 | 0.30 | 0.005 | 2.4 | 0.00051 | 2.5 | 0.00006 | 2.5 |
| 90 | 0.08 | 0.30 | 0.005 | 2.4 | 0.00051 | 2.5 | 0.00006 | 2.5 |
| 120 | 0.08 | 0.30 | 0.005 | 2.4 | 0.00051 | 2.5 | 0.00006 | 2.5 |
| 150 | 0.11 | 0.30 | 0.005 | 2.4 | 0.00051 | 2.5 | 0.00006 | 2.5 |
| 180 | 0.08 | 0.30 | 0.005 | 2.4 | 0.00051 | 2.5 | 0.00006 | 2.5 |

**Table S7:** Fitting Parameters for NSLS-II (large-beam) 100% flux. All $g_2$ functions at different $\phi$ values were co-fit allowing only $\beta$ to independently vary.

| $\phi$ (°) | β | X | q (Å$^{-1}$) | v (Å/s) | $\Gamma_A$ (s$^{-1}$) | $\alpha_A$ | $\Gamma_B$ (s$^{-1}$) | $\alpha_B$ |
|---|---|---|---|---|---|---|---|---|
| 0 | 0.11 | 0.57 | 0.0185 | 2.9 | 0.040 | 0.4 | 0.00020 | 1.5 |
| 30 | 0.10 | 0.57 | 0.0185 | 2.9 | 0.040 | 0.4 | 0.00020 | 1.5 |
| 60 | 0.10 | 0.57 | 0.0185 | 2.9 | 0.040 | 0.4 | 0.00020 | 1.5 |
| 90 | 0.10 | 0.57 | 0.0185 | 2.9 | 0.040 | 0.4 | 0.00020 | 1.5 |
| 120 | 0.09 | 0.57 | 0.0185 | 2.9 | 0.040 | 0.4 | 0.00020 | 1.5 |
| 150 | 0.09 | 0.57 | 0.0185 | 2.9 | 0.040 | 0.4 | 0.00020 | 1.5 |
| 180 | 0.09 | 0.57 | 0.0185 | 2.9 | 0.040 | 0.4 | 0.00020 | 1.5 |

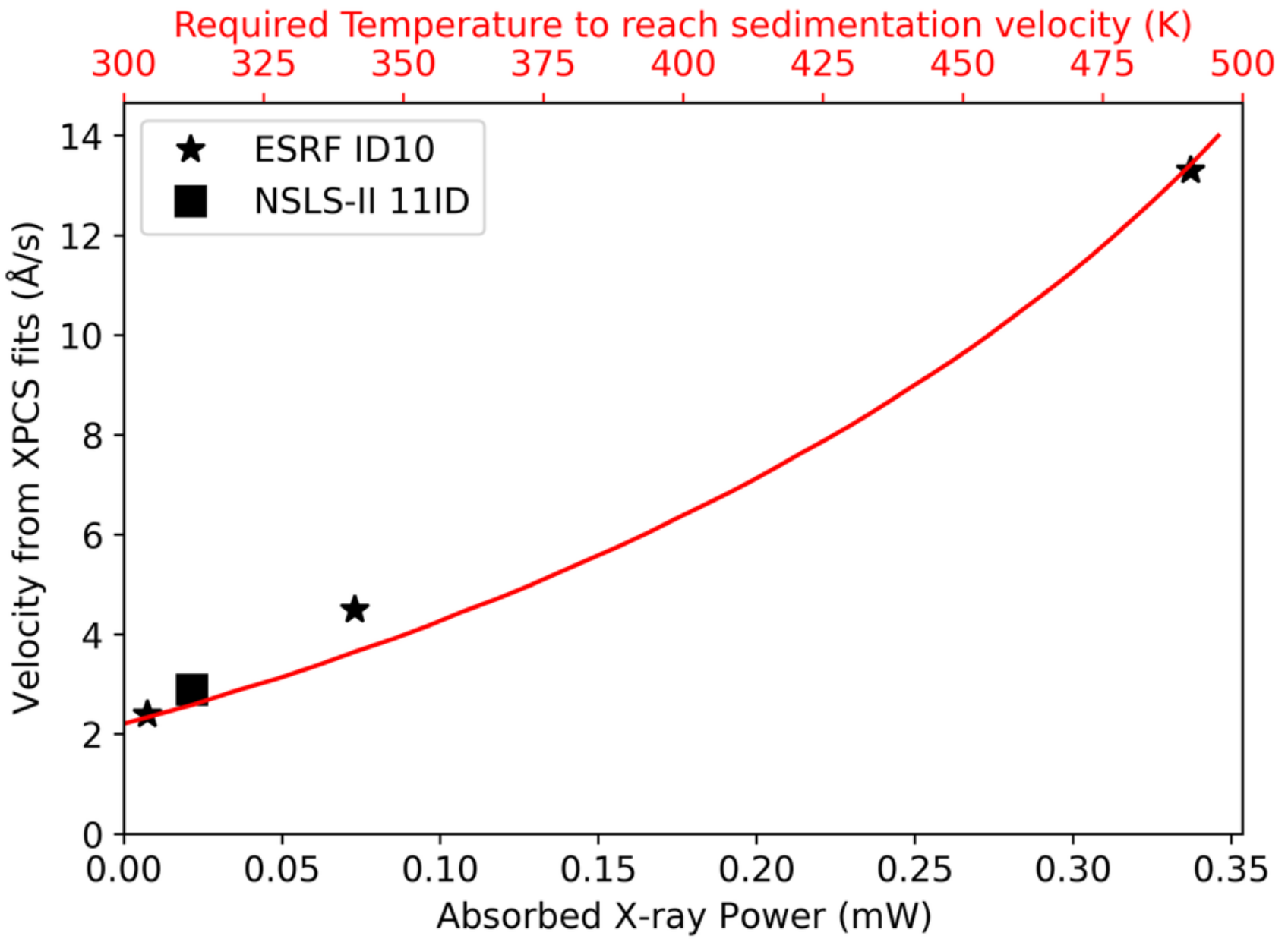

**Figure S12:** Comparison of velocity recorded for experiments of different absorbed X-ray beam power compared with the required solution temperature to reach those velocities as sedimentation in toluene with an assumed polymer aggregate of radius 100 Å. Temperatures were found via Stokes equation assuming temperature dependent viscosity and density of toluene with constant density of polymer aggregate.

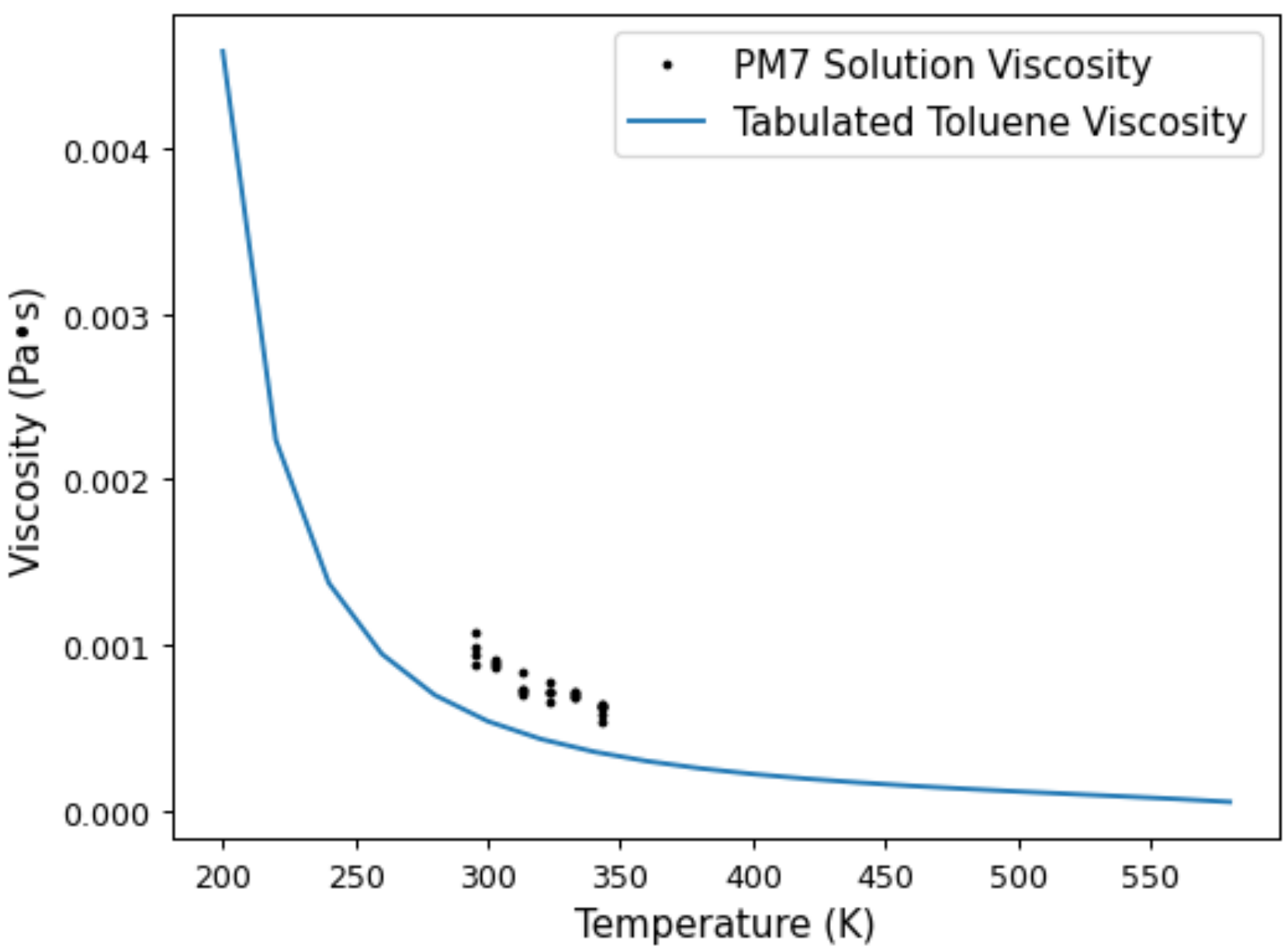

**Figure S13:** Temperature dependent viscosity measurements of PM7-toluene solution compared with neat toluene from literature [2]. Viscosity of PM7 solution is measured at a shear rate of 22,000 $s^{-1}$.

**Comparison of flow velocities from XPCS data fitting and FEA Simulations**

The amount of absorbed power by the solution during XPCS experiments is given by eq. S3.

$$Q = \text{incident flux} \times E_{\text{photon}} \times (1 - e^{-\mu x}) \qquad (S3)$$

Where $\mu$ is the solution absorption coefficient given in the table below and $x$ is 1 mm for all experiments.

COMSOL Multiphysics FEA simulations were used to simulate beam-induced heating and the resulting convective fluid flow. The system was modeled in two dimensions, representing a cross section of a 1 mm diameter capillary with a horizontal length of 10 mm. The X-ray beam was modeled as a 2D gaussian heat source with FWHM equal to the beam dimension on each direction and integrated area equal to $Q$ as shown in equation S3.

The fluid was modeled as incompressible and undergoing laminar flow. We assume the thermal conductivity, heat capacity, thermal expansion, and viscosity of the fluid are the same as neat toluene [3]. The inner walls of the capillary were maintained at a constant temperature of 20 °C simulating the large thermal mass of the metal capillary holder at ambient temperature. The simulation results presented correspond to the steady-state solution for each beam power and geometry. We note that in time-dependent solutions the system rapidly reaches steady state behavior in much less than 1 second.

**Table S8:** Experimental parameters used in the FEA simulations and the extracted values from both FEA simulations and experimental data fitting (Tables S13-S16).

| **Experiment** | **X-ray Energy (J)** | **Beam flux (ph/s)** | **Beam Area ($m^2$)** | **$\mu$ ($mm^{-1}$)** | **Max temp. FEA (K)** | **Max velocity FEA (m/s)** | **Flow velocity XPCS (m/s)** |
|---|---|---|---|---|---|---|---|
| ESRF: 47% flux | $1.6 \times 10^{-15}$ | $1.2 \times 10^{12}$ | $4.0 \times 10^{-11}$ | 0.20 | 295.6 K | $1.3 \times 10^{-4}$ | $1.3 \times 10^{-9}$ |
| ESRF: 9% flux | $1.6 \times 10^{-15}$ | $2.3 \times 10^{11}$ | $4.0 \times 10^{-11}$ | 0.20 | 293.7 K | $2.7 \times 10^{-5}$ | $4.5 \times 10^{-10}$ |
| ESRF: 1% flux | $1.6 \times 10^{-15}$ | $2.5 \times 10^{10}$ | $4.0 \times 10^{-11}$ | 0.20 | 293.2 K | $2.7 \times 10^{-6}$ | $2.4 \times 10^{-10}$ |
| NSLS: 100% flux | $1.9 \times 10^{-15}$ | $1.0 \times 10^{11}$ | $1.6 \times 10^{-9}$ | 0.12 | 293.3 K | $1.1 \times 10^{-5}$ | $2.9 \times 10^{-10}$ |

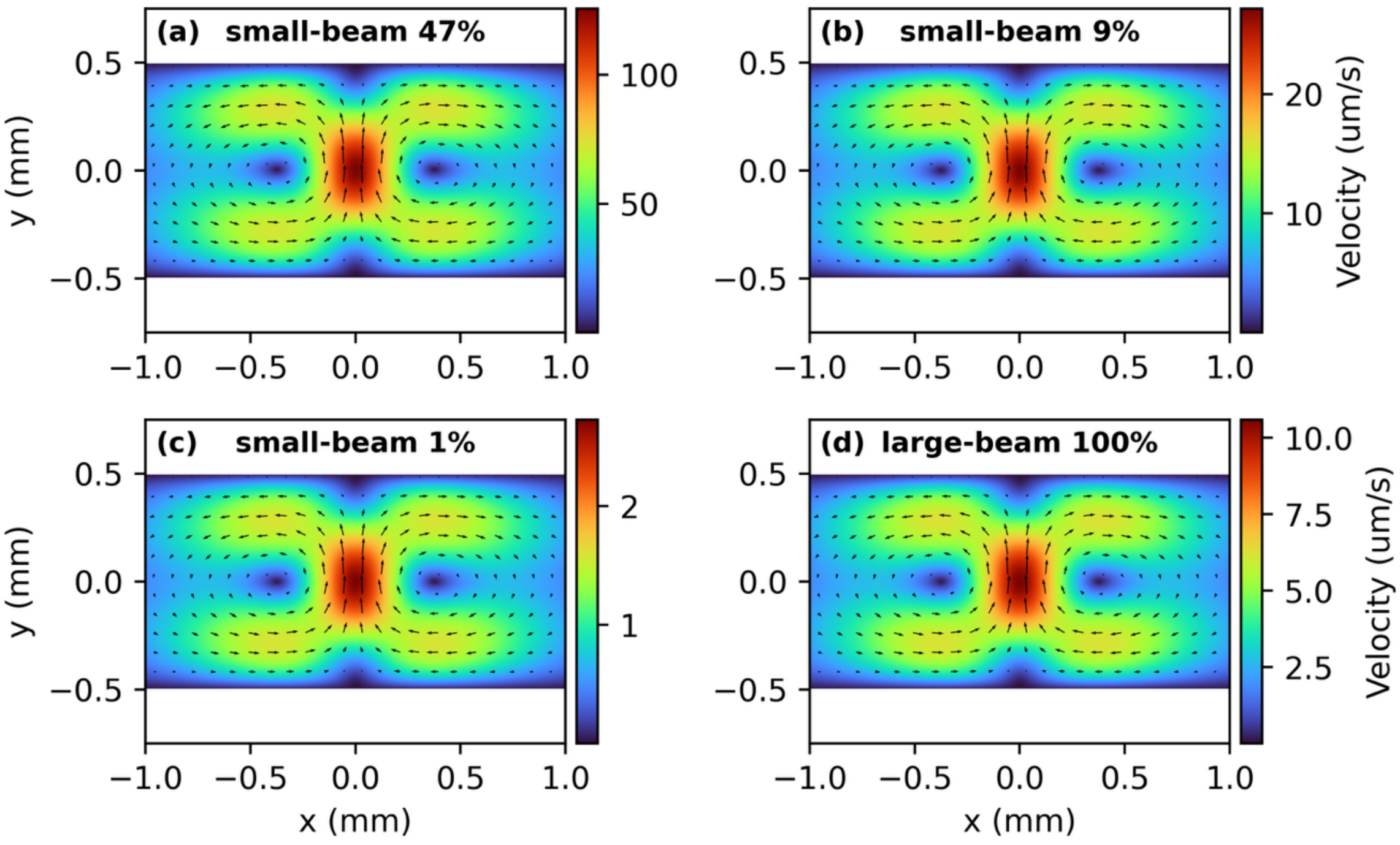

**Figure S14**: flow map from FEA simulations of (a) small-beam experiment with 47% flux, (b) small-beam experiment with 9% flux, (c) small-beam experiment with 1% flux, (d) large-beam experiment with 100% flux.

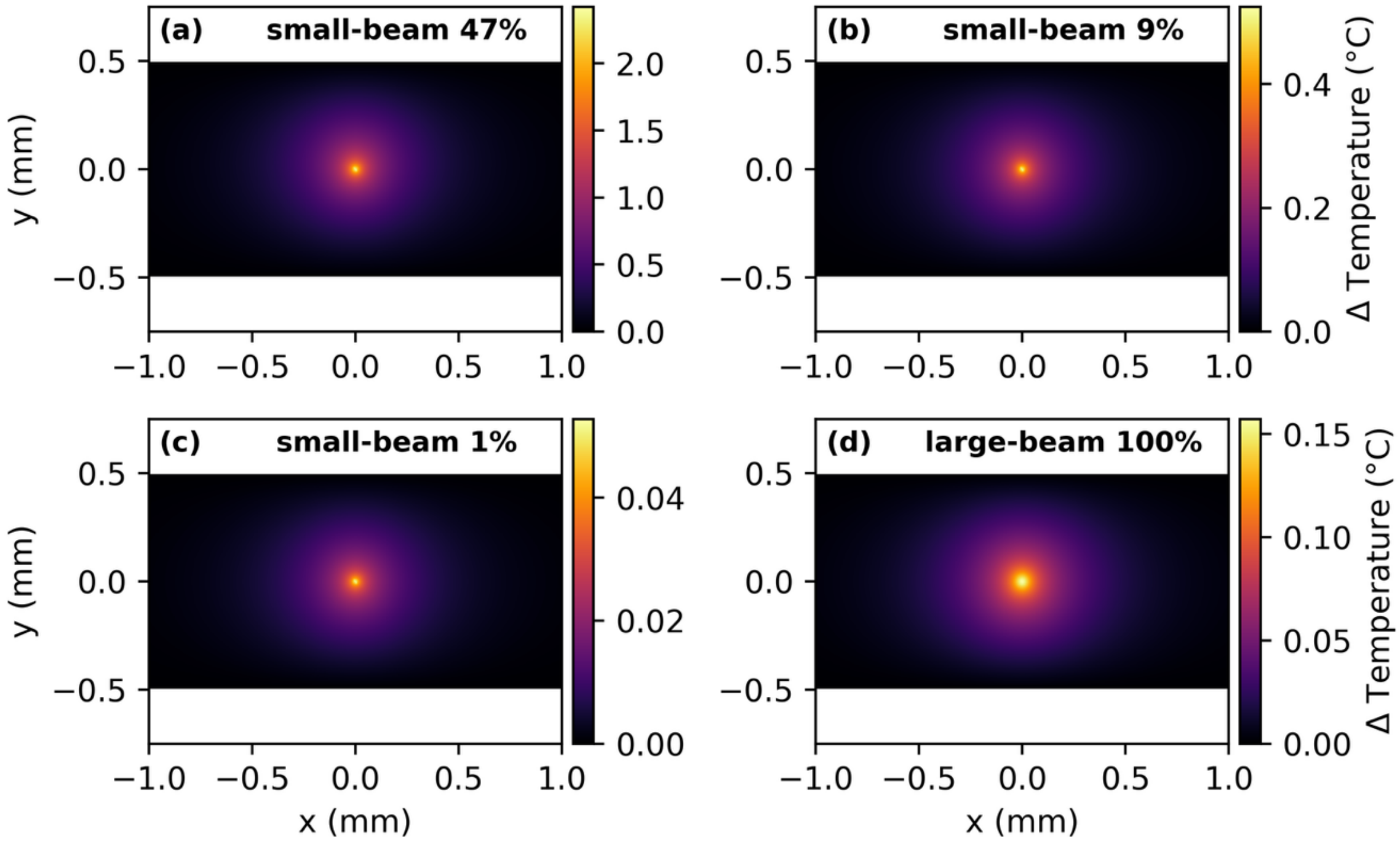

**Figure S15**: temperature maps from FEA simulations showing the rise in temperature compared to ambient 20 °C during (a) small-beam experiment with 47% flux, (b) small-beam experiment with 9% flux, (c) small-beam experiment with 1% flux, (d) large-beam experiment with 100% flux.